\documentclass[12pt]{article}
\usepackage{graphicx}
\usepackage{amsmath}
\usepackage{hyperref}
\usepackage{tabularx,booktabs}
\usepackage{enumitem}
\usepackage{multirow}
\usepackage{array}
\usepackage{enumitem}
\usepackage{nameref}
\usepackage{adjustbox}
\usepackage{xcolor}
\usepackage{tikz}
\usepackage{subcaption}
\usetikzlibrary{positioning}
\usepackage{tikz}

\hypersetup{
    colorlinks=true,
    linkcolor=blue,      
    urlcolor=blue,       
    citecolor=blue       
}

\newlist{researchq}{description}{1}
\setlist[researchq,1]{labelwidth=\widthof{\bfseries RQ\ref{rqi}}, leftmargin=!}

\newcolumntype{P}[1]{>{\raggedright\arraybackslash}p{#1}}
\newcolumntype{M}[1]{>{\raggedright\arraybackslash\ttfamily}m{#1}}
\newcommand{\todo}[1]{}
\renewcommand{\todo}[1]{\pdfliteral{1 0 0 rg}#1\pdfliteral{0 0 0 rg}}

\makeatletter
\let\orgdescriptionlabel\descriptionlabel
\renewcommand*{\descriptionlabel}[1]{%
	\let\orglabel\label
	\let\label\@gobble
	\phantomsection
	\protected@edef\@currentlabel{#1\unskip}%
	\let\label\orglabel
	\orgdescriptionlabel{#1}%
}
\makeatother

\begin{document}

\title{2D Matryoshka Training for Information Retrieval}

		
	%
	%

\author{
	Shuai Wang\thanks{Equal contribution} \\
	The University of Queensland \\
	\texttt{shuai.wang2@uq.edu.au}
	\and
	Shengyao Zhuang\footnotemark[1] \\
	CSIRO \\
	\texttt{shengyao.zhuang@csiro.com}
	\and
	Bevan Koopman \\
	CSIRO \\
	\texttt{b.koopman@csiro.com}
	\and
	Guido Zuccon \\
	The University of Queensland \\
	\texttt{g.zuccon@uq.edu.au}
}
\maketitle

\begin{abstract}
	2D Matryoshka Training is an advanced embedding representation training approach designed to train an encoder model simultaneously across various layer-dimension setups. This method has demonstrated higher effectiveness in Semantic Text Similarity (STS) tasks over traditional training approaches when using sub-layers for embeddings. Despite its success, discrepancies exist between two published implementations, leading to varied comparative results with baseline models. In this reproducibility study, we implement and evaluate both versions of 2D Matryoshka Training on STS tasks and extend our analysis to retrieval tasks. Our findings indicate that while both versions achieve higher effectiveness than traditional Matryoshka training on sub-dimensions, and traditional full-sized model training approaches, they do not outperform models trained separately on specific sub-layer and sub-dimension setups. Moreover, these results generalize well to retrieval tasks, both in supervised (MSMARCO) and zero-shot (BEIR) settings. Further explorations of different loss computations reveals more suitable implementations for retrieval tasks, such as incorporating full-dimension loss and training on a broader range of target dimensions. Conversely, some intuitive approaches, such as fixing document encoders to full model outputs, do not yield improvements. Our reproduction code is available at \url{https://github.com/ielab/2DMSE-Reproduce}.
	
\end{abstract}

\section{Introduction}

Transformer-based dense retrievers operate by encoding queries and documents into dense vectors in separate processes: document vectors are pre-computed and stored during an offline phase, while query encoding and subsequent vector comparisons occur online when a query is issued to the retrieval system. This architecture ensures efficiency in ranking by shifting the computationally intensive task of document encoding to an offline stage. Consequently, during the online retrieval phase, the system focuses solely on query processing and the rapid comparison of query vectors against the pre-encoded document vectors. This separation not only speeds up the retrieval process but also allows for scaling to large document corpora, making transformer-based dense retrievers particularly effective in environments where quick response times are critical.
However, backbone models used in these retrievers have grown significantly in size to achieve more effective ranking—some now exceeding 7 billion parameters~\cite{repllama}. As a result, these models require significant computational resources during the retrieval for the query encoding phase alone~\cite{repllama}.

The 2D Matryoshka sentence embedding training (2DMSE)\footnote{For simplicity, we refer to this as 2D Matryoshka training.} is a recent innovation that introduces a unique training strategy that trains embeddings from each layer of a model simultaneously across different dimension sizes~\cite{li2024eseespressosentenceembeddings}. This allows for the extraction of embeddings directly from sub-layers rather than processing through the entire model, significantly reducing the time required for encoding text. This approach has shown promising results, with embeddings trained using 2D Matryoshka training demonstrating higher effectiveness than those from a standard full-size model in Semantic Text Similarity (STS) tasks.

While 2DMSE intuitively applies to information retrieval tasks, due to its ability to generate embeddings similar to those used in dense retrievers, the nature of these tasks differs. In STS, the focus is on closely aligning the semantic meanings of two sentences, typically of similar length and structure. In contrast, information retrieval involves matching queries, often brief, to longer documents, which presents unique challenges due to the disparity in text length and content presentation. Additionally, the training methodologies for dense retrievers diverge from those in STS. Instead of merely aligning semantic meanings, training for retrieval tasks emphasizes optimizing the relevance of a set of documents in response to specific queries. 

In this reproducibility study, our objective extends beyond merely reproducing the original results of 2DMSE on STS tasks. We also aim to examine its adaptability and effectiveness within the domain of information retrieval. To comprehensively assess its applicability and performance in this new context, we have formulated the following four research questions:

\begin{description}[leftmargin=4pt]
	\item[RQ1] \textit{Can we replicate 2D Matryoshka training for Semantic Text Similarity?} 	
	The original study shows that 2DMSE outperforms both Matryoshka training and full model training, especially when fewer layers are used for embeddings~\cite{li2024eseespressosentenceembeddings}. 2DMSE also surpasses separately trained sub-models in smaller layer and dimension configurations. However, the original 2D Matryoshka paper includes two design versions with varying implementation details. This study aims to: (1) Verify whether the results are consistent across both versions of the original paper, and (2) Expand the comparison with separately trained small models to encompass more model cases, rather than the limited scenarios\footnote{Only two sub-models were tested in the original work.} previously considered, to provide a comprehensive analysis of performance differentials.
	
	\item[RQ2] \textit{Does 2D Matryoshka training generalize to information retrieval tasks?} We evaluate the effectiveness on information retrieval, analyzing whether the advantages observed in STS extend to this new setting.
	
	\item[RQ3] \textit{Does a modified loss calculation in 2DMSE improve retrieval effectiveness?} The original implementation of 2DMSE did not consider the unique characteristics of retrieval tasks, such as using embeddings for KL divergence, instead of learning directly from output scores. We then investigate whether modifying the loss computation can enhance retrieval effectiveness.
	
	\item[RQ4] 
	\textit{Can a small, 2DMSE query encoder be used in conjunction with a full document encoder to improve effectiveness and efficiency in retrieval?} In dense retrieval, documents are typically encoded offline, so that document processing does not affect online query latency, which is critical in search scenarios. We explore whether employing a full model for document encoding, while pruning the query encoder, can enhance the overall effectiveness of the model without compromising online query latency.
	
\end{description}

Our reproducibility study makes several key contributions. First, we \textbf{reproduce} and \textbf{validate} the original findings from the 2D Matryoshka paper on the Semantic Textual Similarity (STS) task, ensuring the reliability of the reported results. Second, we extend the application of the 2D Matryoska model to the domain of dense retrieval, investigating its \textbf{transferability} and effectiveness in a new context. Third, we delve into the optimal methods of applying the 2D Matryoska model specifically for retrieval tasks, incorporating strategies tailored to enhance retrieval effectiveness. Finally, we commit to open science principles by publishing reproducible and well-documented code for implementing the 2D Matryoska model on both STS and dense retrieval tasks.


\section{2D Matryoshka Training (2DMSE)}

2D Matryoshka training~\cite{li2024eseespressosentenceembeddings} features two distinct design and implementation versions, each described in two separate versions of the original arXiv publication. This section provides an overview of both designs. We begin with a discussion on Matryoshka training in Section~\ref{sec:1dmse}. Subsequently, the two versions of 2DMSE training are detailed in Section~\ref{sec:2dmse_v1} and Section~\ref{sec:2dmse_v2}, respectively.

\subsection{Matryoshka Training}
\label{sec:1dmse}

Matryoshka training~\footnote{Also known as Matryoshka Representation Learning} is inspired by the hierarchical structure of Matryoshka dolls, where progressively smaller dolls are encased within larger ones. Similarly, in the domain of embedding training, this approach involves the simultaneous learning of multiple embeddings, each with differing dimension sizes~\cite{mrl2022}.

To train Matryoshka representations, an encoder model \(\mathcal{E}\), typically acting as the backbone of the approach, is tasked with multiple training objectives derived from the final layer's various dimensional outputs. The process can be formalized as follows: Given an input \(\mathbf{x}\), the encoder \(\mathcal{E}\) generates a ``full'' embedding \(\mathbf{e}^{(L)}\), which comprises the complete set of dimensions (from the last layer \(L\)):
\begin{equation}
	\mathbf{e}^{(L)} = \mathcal{E}(\mathbf{x}; \Theta)
\end{equation}
where \(\Theta\) denotes the parameters of the encoder. From this comprehensive embedding, a subset of dimensions is selected to form reduced-dimension embeddings \(\{\mathbf{e}^{(L)}_{k_1}, \mathbf{e}^{(L)}_{k_2}, \dots, \mathbf{e}^{(L)}_{k_n}\}\), where each \(\mathbf{e}^{(L)}_{k_i}\) represents the first \(k_i\) dimensions of \(\mathbf{e}^{(L)}\):

\begin{equation}
	\mathbf{e}^{(L)}_{k_i} = \mathbf{e}^{(L)}[:k_i], \quad i = 1, 2, \dots, n
\end{equation}

The objective is to minimize the composite loss function, which is a summation of the losses associated with each dimension-specific embedding:

\begin{equation}
	\mathcal{L}(\Theta) = \sum_{i=1}^n \ell(\mathbf{e}^{(L)}_{k_i}, \mathbf{y})
\end{equation}
\(\ell\) represents the loss function for each embedding, and \(\mathbf{y}\) is the target output. The loss function can be any defined loss function, such as contrastive loss~\cite{carlssonsemantic,wang2021understanding}.

\subsection{2D Matryoshka Training Version 1 (2DMSE-V1) }
\label{sec:2dmse_v1}

The first version of the 2D Matryoshka training (2DMSE-V1) utilizes embeddings from both the last layer and a randomly selected sub-layer, incorporating Kullback-Leibler (KL) divergence to align the distribution of these embeddings. This approach trains embeddings from the last layer simultaneously to those from previous layers. The detailed steps of 2DMSE-v1 are as follows:

\begin{enumerate}

	\item \textbf{Last Layer Embedding:} The embeddings from the last layer, denoted as \(\mathbf{e}^{(L)}\), are responsible to capture the most comprehensive and detailed representations of the input data. The loss associated with these embeddings is defined as:
	\begin{equation}
		\mathcal{L}_{\text{last}}(\Theta) = \ell(\mathbf{e}^{(L)}, \mathbf{y})
	\end{equation}

	\item \textbf{Sub-Layer Embedding:} Simultaneously, embeddings from a randomly selected sub-layer (earlier layer) \(r\) (where \(1 \leq r < L\)) are extracted, denoted as \(\mathbf{e}^{(r)}\), to provide insights from a different level of abstraction of the input. The loss for these embeddings is given by:
	\begin{equation}
		\mathcal{L}_{\text{random}}(\Theta) = \ell(\mathbf{e}^{(r)}, \mathbf{y})
	\end{equation}

	\item \textbf{KL Divergence Loss:} To ensure that the embeddings maintain a consistent distribution regardless of the layer they are sourced from, KL divergence is employed to measure the distance between the probability distributions of the last layer's embeddings and the randomly selected layer's embeddings:
	
	\begin{equation}
		\mathcal{L}_{\text{KLD}}(\Theta) = \text{KLD}(\mathbf{p}(\mathbf{e}^{(L)}) \parallel \mathbf{p}(\mathbf{e}^{(r)}))
	\end{equation}
\end{enumerate}

The total loss function combines these individual components, ensuring both accuracy and consistency in the embedding process:

\begin{equation}
	\mathcal{L}(\Theta) = \mathcal{L}_{\text{last}}(\Theta) + \mathcal{L}_{\text{random}}(\Theta) + \lambda \mathcal{L}_{\text{KLD}}(\Theta)
\end{equation}

Here, \(\lambda\) is a scaling factor regulating the impact of the KL divergence on the total loss, \(\ell\) denotes the loss function for each set of embeddings, \(\text{KLD}\) is the Kullback–Leibler divergence, and \(\mathbf{p}(\mathbf{e})\) indicates the probability distribution of embeddings \(\mathbf{e}\). This method fosters a nuanced and coherent learning strategy across different layers of the encoder.

\subsection{2D Matryoshka Training Version 2 (2DMSE-V2)}
\label{sec:2dmse_v2}

The second version of 2D Matryoshka training (2DMSE-V2) enhances the model's feature extraction capabilities through tailored modifications to the layer-specific and PCA alignment losses:

\begin{enumerate}

	\item \textbf{Layer-specific Loss:} All layers of the encoder are utilized, with a logarithmic weighting function applied to all but the last layer, which receives a weight of 1. This differential weighting is specified as:
	
	\begin{equation}
		w_i = \left\{
		\begin{array}{ll}
			\dfrac{1}{1 + \ln(i)} & \text{for } i < L \\
			1 & \text{for } i = L
		\end{array}
		\right.
	\end{equation}
	
	The layer-specific loss is then calculated by applying these weights to the losses associated with the embeddings from each layer, particularly focusing on the specified sub-dimensions \( k_i \):
	
	\begin{equation}
		\mathcal{L}_{\text{layer}}(\Theta) = \sum_{i=1}^{L} w_i \cdot \ell(\mathbf{e}^{(i)}_{k_i}, \mathbf{y}_{k_i})
		\label{equ:9}
	\end{equation}

	\item \textbf{Dimension-specific Loss:} The dimension-specific loss employs the same weighted structure as the layer-specific losses. For each layer \(i\), the embeddings \(\mathbf{e}^{(i)}\) are transformed via PCA and truncated to a target dimension \(k_i\), resulting in \(\tilde{\mathbf{e}}^{(i)}_{k_i}\). Similarly, the target outputs \(\mathbf{y}\) are transformed or truncated to match the dimension \(k_i\), denoted as \(\tilde{\mathbf{y}}_{k_i}\). The PCA-transformed and truncated embeddings across all layers are optimized using a combination of Mean Squared Error (MSE) and KL divergence losses:
	
	\begin{equation}
		\mathcal{L}_{\text{dim}}(\Theta) = \sum_{i=1}^{L} w_i \cdot \left( \text{MSE}(\tilde{\mathbf{e}}^{(i)}_{k_i}, \tilde{\mathbf{y}}_{k_i}) + \text{KLD}(\mathbf{p}(\tilde{\mathbf{e}}^{(i)}_{k_i}) \parallel \mathbf{p}(\tilde{\mathbf{y}}_{k_i})) \right)
		\label{equ:10}
	\end{equation}
\end{enumerate}

The overall loss function integrates these components to maximize the representational power and alignment of the embeddings:

\begin{equation}
	\mathcal{L}(\Theta) = \alpha \mathcal{L}_{\text{layer}}(\Theta) + \beta \mathcal{L}_{\text{dim}}(\Theta)
\end{equation}

\noindent where \(\alpha\) and \(\beta\) are set as 1 by default. This approach ensures that each layer's sub-embedding contributes optimally to the model's performance, and that the full embeddings are aligned in the lower dimensional space through PCA. 

However, upon reviewing the implementation details in the original codebase, we noted that the loss from the full dimension size of the last layer's embeddings is also included in the final loss computation. This observation leads to the following adjustment in the total loss function:

\begin{equation} 
	\mathcal{L}(\Theta) = \alpha \mathcal{L}{\text{layer}}(\Theta) + \beta \mathcal{L}{\text{dim}}(\Theta) + \ell(\mathbf{e}^{(L)}, \mathbf{y}) 
\end{equation}

Additionally, a critical observation was made concerning the training dependencies described in the original study (both the paper and its implementation): the full dimension embeddings from the sub-layers were never directly trained; only the target sub-dimensions were explicitly optimized, as detailed in Equation~\ref{equ:9}. This suggests that the truncated dimension should be considered the primary guide or 'teacher' for the learning of the full-dimension embeddings, which contradicts the initial interpretation in the paper that posited the full-dimension setup as the teacher. In our study, we explore the implications of also training the full-dimension sub-layers and demonstrate the outcomes in Section~\ref{sec:rq3}.



\section{Experimental Setup}

This section outlines the experimental framework, focusing on two primary tasks: Semantic Text Similarity (STS) and Passage Retrieval. We use \texttt{bert-base-} \texttt{uncased} as the initial model for tuning across all experiments. The subsections below describe the training and evaluation methodologies, datasets employed, and configurations used for each task. For Version 2 (V2), we use target embedding \(k_i = 128\) -- the same configuration as the original study. Our baseline includes (1) full models trained using last layer embeddings at full dimensions, (2) sub-models trained on targeted evaluated layer and dimension setups, and (3) Matryoshka-trained models focusing on sub-dimensions from the last layers.

\subsection{Semantic Text Similarity (STS)}

\paragraph{\bf Model Training:} We replicate the training setup from the original study by using the All-NLI dataset, which amalgamates the Multi-Genre Natural Language Inference (MultiNLI)~\cite{williams-etal-2018-broad} and the Stanford Natural Language Inference (SNLI)~\cite{bowman-etal-2015-large} datasets. Specifically, we use the Huggingface dataset \href{https://huggingface.co/datasets/SeanLee97/nli_for_simcse}{SeanLee97/nli\_for\_simcse} to formulate training samples.

Training is conducted under two distinct 2D Matryoshka representation learning frameworks:

\begin{itemize}

	\item \textbf{2DMSE-V1:} Utilizes the \texttt{sentence\_transformers} library~\cite{reimers-gurevych-2019-sentence}, incorporating the initial implementation of 2D Matryoshka embeddings with a MultipleNegativeLoss function for training.
	\item \textbf{2DMSE-V2:} Utilizes the AnglE library~\cite{li-li-2024-aoe}, which facilitates specific training procedures tailored for the second version of the framework with an Angle loss function.
\end{itemize}

For standard hyperparameters, we employ example BERT fine-tuning parameters from the AnglE library, with detailed parameter settings listed in our GitHub repository.\footnote{\url{https://github.com/ielab/2DMSE-Reproduce/}} We use BERT training parameters as the original 2DMSE-V2 configurations have not been reported.\footnote{Refer to issue~\url{https://github.com/SeanLee97/AnglE/issues/80}}

\textbf{Evaluation:}
Model effectiveness is evaluated with several Semantic Text Similarity benchmarks from SemEval 2012--2016~\cite{agirre-etal-2016-semeval}, the SICK-Relatedness dataset (SICK-R)~\cite{marelli-etal-2014-sick}, and the STS Benchmark (STS-B)~\cite{cer-etal-2017-semeval}. We use Spearman's rank correlation coefficient as the evaluation metric as per common practice for this task. To manage computational demands, we evaluate a subset of models with six layers $[2, 4, 6, 8, 10, 12]$ (instead of all layers) and eight dimensional settings $[8, 16, 32, 64, 128, 256, 512, 768]$ (as in the initial 2DMSE work), totalling to 48 different combinations.

For a fair comparison, V2 models are converted to a format compatible with the \texttt{sentence\_transformers} library and assessed using a standardized script.

\subsection{Passage Retrieval}

\paragraph{\bf Model Training:}

We adopt contrastive loss for fine-tuning, utilizing the Tevatron library~\cite{tevatron} for training our retrieval models; this is done on the MS MARCO Passage Ranking dataset~\cite{bajaj2018msmarcohumangenerated}. The standard training setup from Tevatron is followed, with fine-tuning parameters detailed in our GitHub repository. 

\paragraph{\bf Evaluation:} For supervised setting, models are evaluated on the MS MARCO development set~\cite{bajaj2018msmarcohumangenerated} using MRR@10. 
For zero-shot setting, models are evaluated using the BEIR benchmark~\cite{beir}, focusing on evaluation across diverse text retrieval tasks and domains. We use ndcg@10 as our target evaluation metric on 13 publicly available BEIR datasets. 
We similarly examine 48 models based on a matrix of six layers and eight dimension configurations.



\section{Results and Analysis}

\begin{figure}
	\centering
	\begin{subfigure}[t]{0.51\textwidth}
		\centering
		\includegraphics[width=\textwidth]{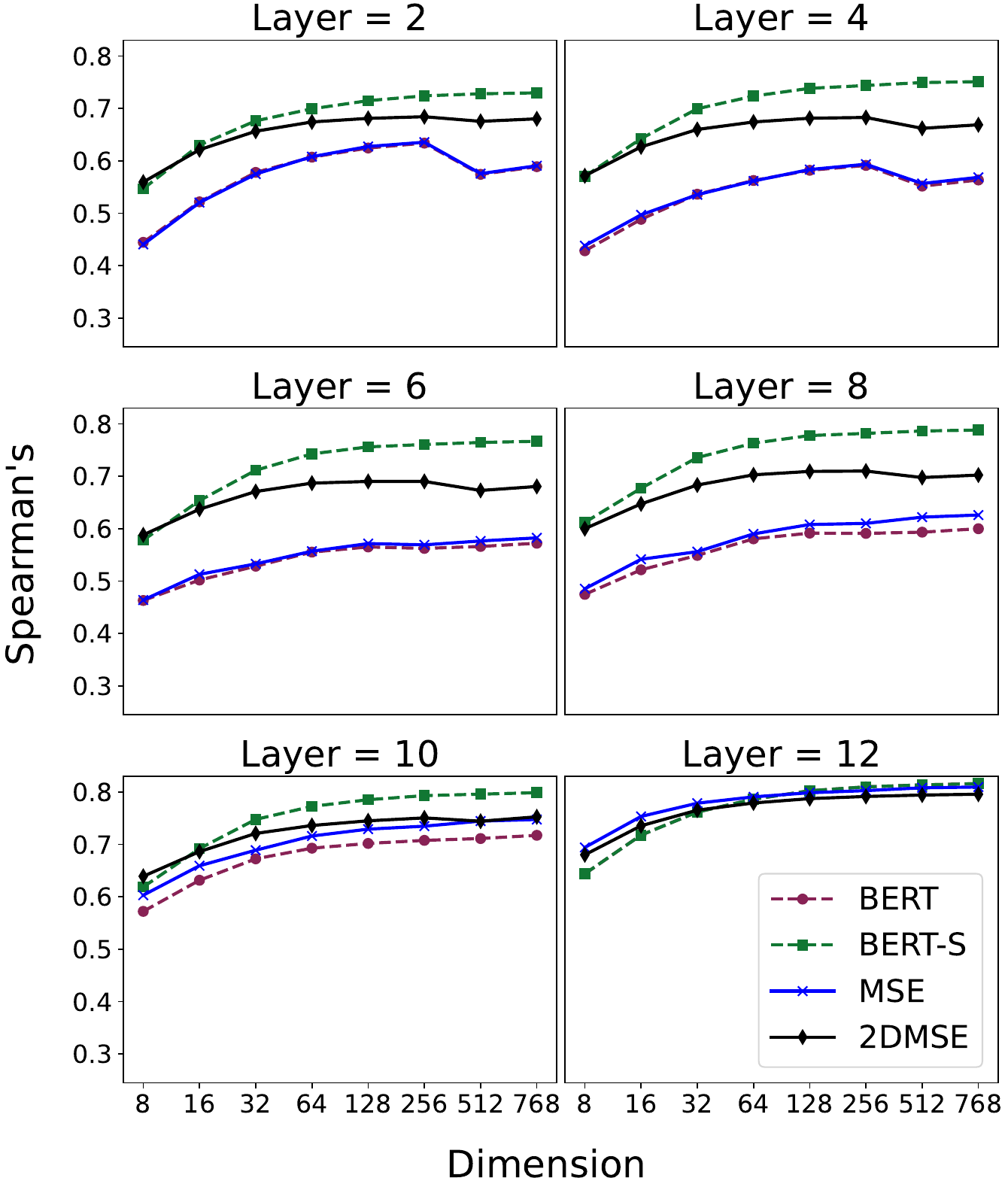}
		\caption{2DMSE-V1}
		\label{fig:sts_v1}
	\end{subfigure}
	\hfill
	\begin{subfigure}[t]{0.48\textwidth}
		\centering
		\includegraphics[width=\textwidth]{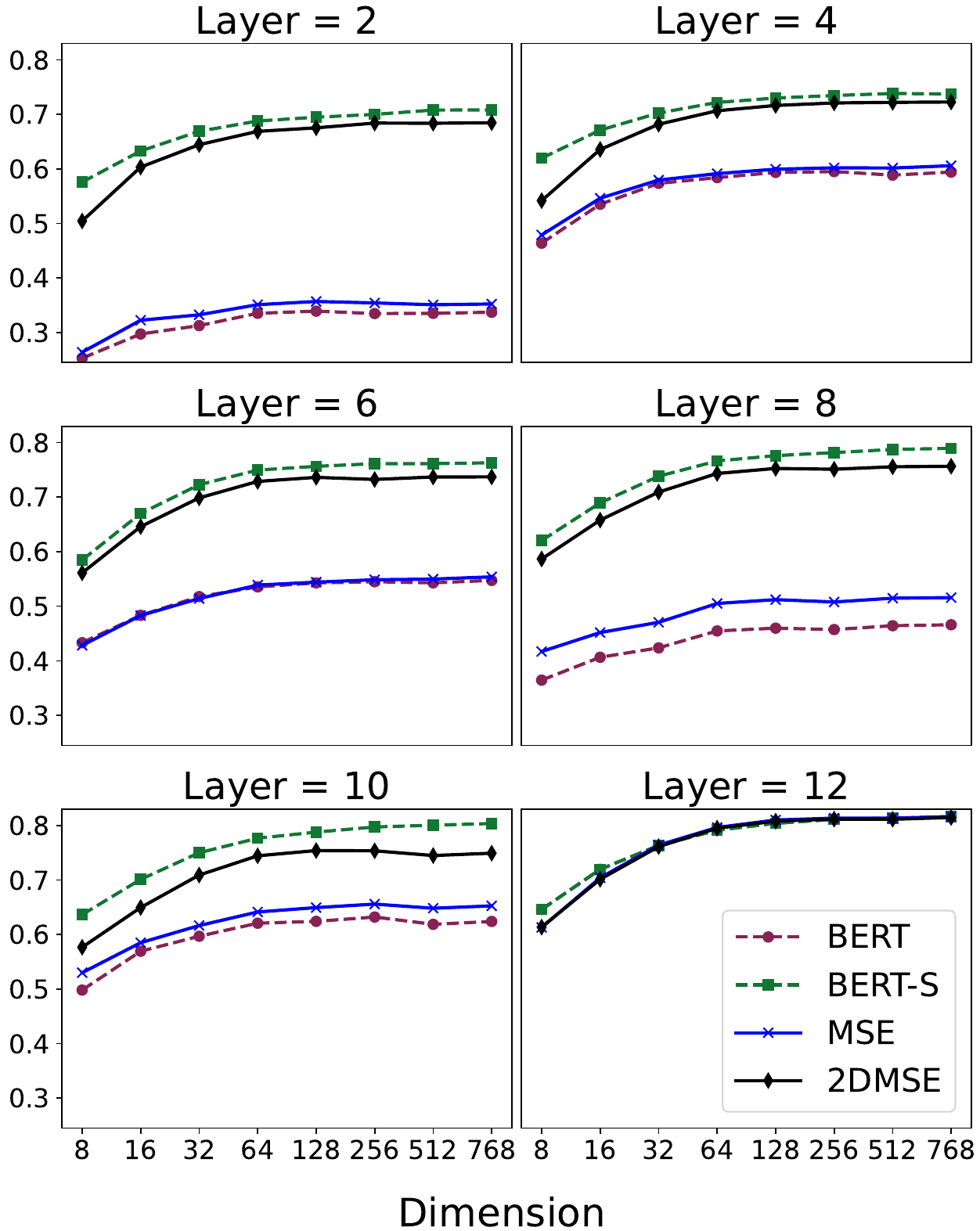}
		\caption{2DMSE-V2}
		\label{fig:sts_v2}
	\end{subfigure}
	\caption{Comparison of model effectiveness across different layer and dimension setups for the STS task, evaluated using Spearman's Correlation.}
	\label{fig:sts_main}
\end{figure}

In this section, we present the results of our reproducibility study.
All results for the STS task are shown in Figure~\ref{fig:sts_main}, which reports Spearman's correlation averages for the baseline and MSE models across various layer-dimension setups~\footnote{Full tabulated results at~\href{https://github.com/ielab/2DMSE-Reproduce/}{Github Repository}.}.
All results for the retrieval task are shown in Figure~\ref{fig:retreival_base}, which reports MRR@10 for the MSMARCO dataset. 
In both figures, \textbf{BERT} denotes the fine-tuning of a BERT model with loss only from the last layer and full dimension, subsequently applied to different layer-dimension setups for evaluation.\footnote{This corresponds to the RAW configuration in the original work.} \textbf{BERT-S} represents models separately fine-tuned for each layer and dimension setup. \textbf{2DMSE} refers to the use of a 2D Matryoshka training strategy to train a single model, evaluated across different setups. \textbf{MSE} involves using a Matryoshka training strategy to fine-tune the model on the last layer with various dimension setups (with a similar loss structure as in 2DMSE, but computed only for the last layer).

\subsection{RQ1: Can we replicate 2D Matryoshka training for Semantic Text Similarity?}

The original study concluded that: (1) the 2DMSE-trained model achieves higher effectiveness on the sub-layers compared to MSE and BERT; (2) in smaller-scale models, 2DMSE shows greater effectiveness than BERT-S. As the original paper did not provide specific training details or hyperparameter configurations, our goal is not to replicate the absolute metrics reported but rather to confirm the observed trends.

Our results, summarized in Figure~\ref{fig:sts_main}, confirm the original work's findings: the 2DMSE model consistently outperforms MSE and BERT in the sub-layers for all evaluated layers and dimensions. However, with respect to the second conclusion from the original study, we observe that while the first version of 2DMSE surpasses BERT-S in the sub-layers (i.e., layers 2, 4, and 6) and at an embedding size of 8, this is not consistent in the second version of the model. Here, 2DMSE is underperformed by BERT-S, except in the last layer and at dimensions 64, 128, and 256, albeit by a small margin.

When we compare the effectiveness of 2DMSE from the two versions, the plot reveals that in most cases, 2DMSE-V2 achieves a higher effectiveness compared to 2DMSE-V1, and shows a smaller gap compared to separately trained small models. This suggests that 2DMSE-V2 is potentially more suitable for generating high-quality embeddings for STS tasks.

%

\subsection{\textbf{RQ2: Does 2D Matryoshka training generalize to information retrieval tasks?}}

\begin{figure}[t]
	\centering
	\resizebox{\textwidth}{!}{
		\includegraphics[width=\textwidth]{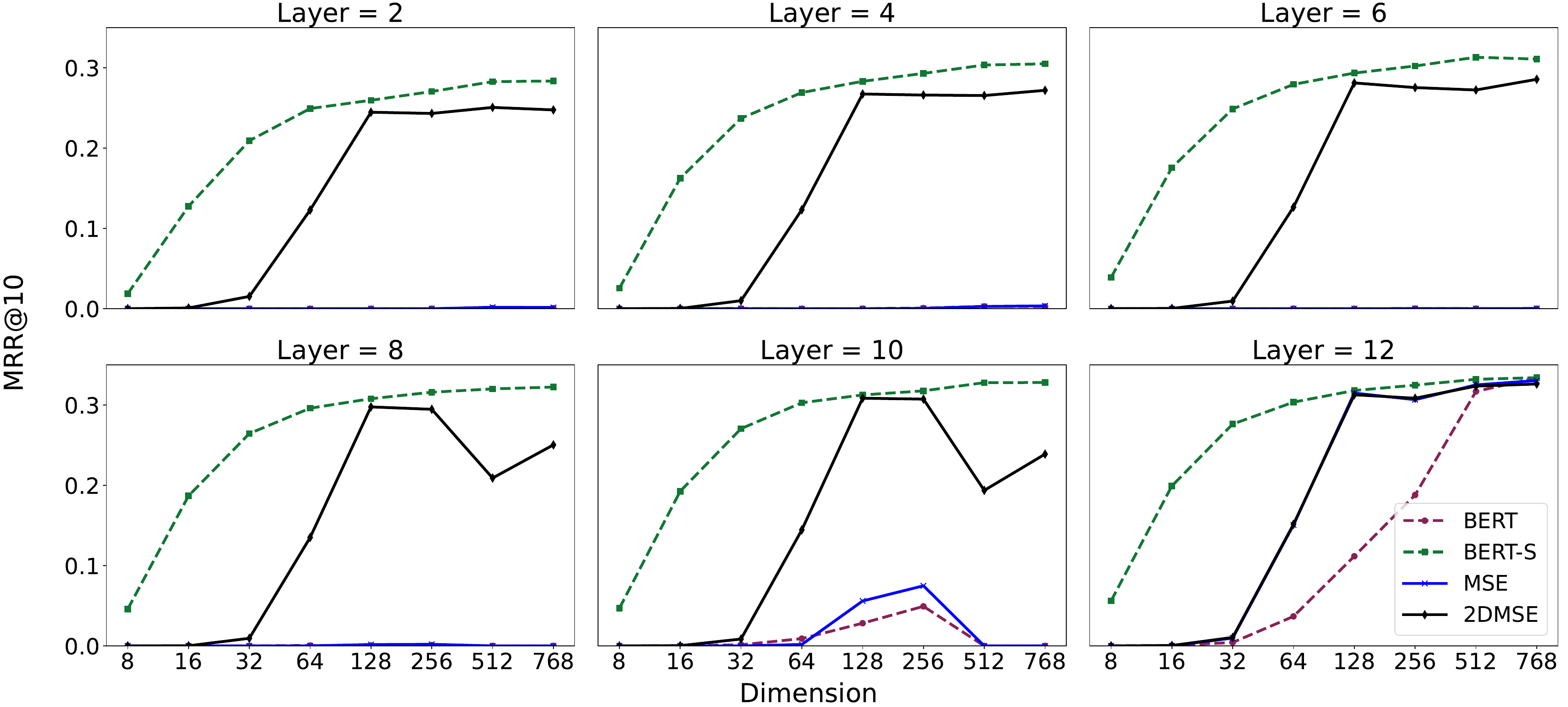}
	}
	\caption{Comparison of Model Effectiveness in different layer and dimension setup for retrieval, evaluated using MRR@10 on MSMARCO.}
	\label{fig:retreival_base}
\end{figure}
\begin{figure}[t]
	\centering
	\resizebox{\textwidth}{!}{
		\includegraphics[width=\textwidth]{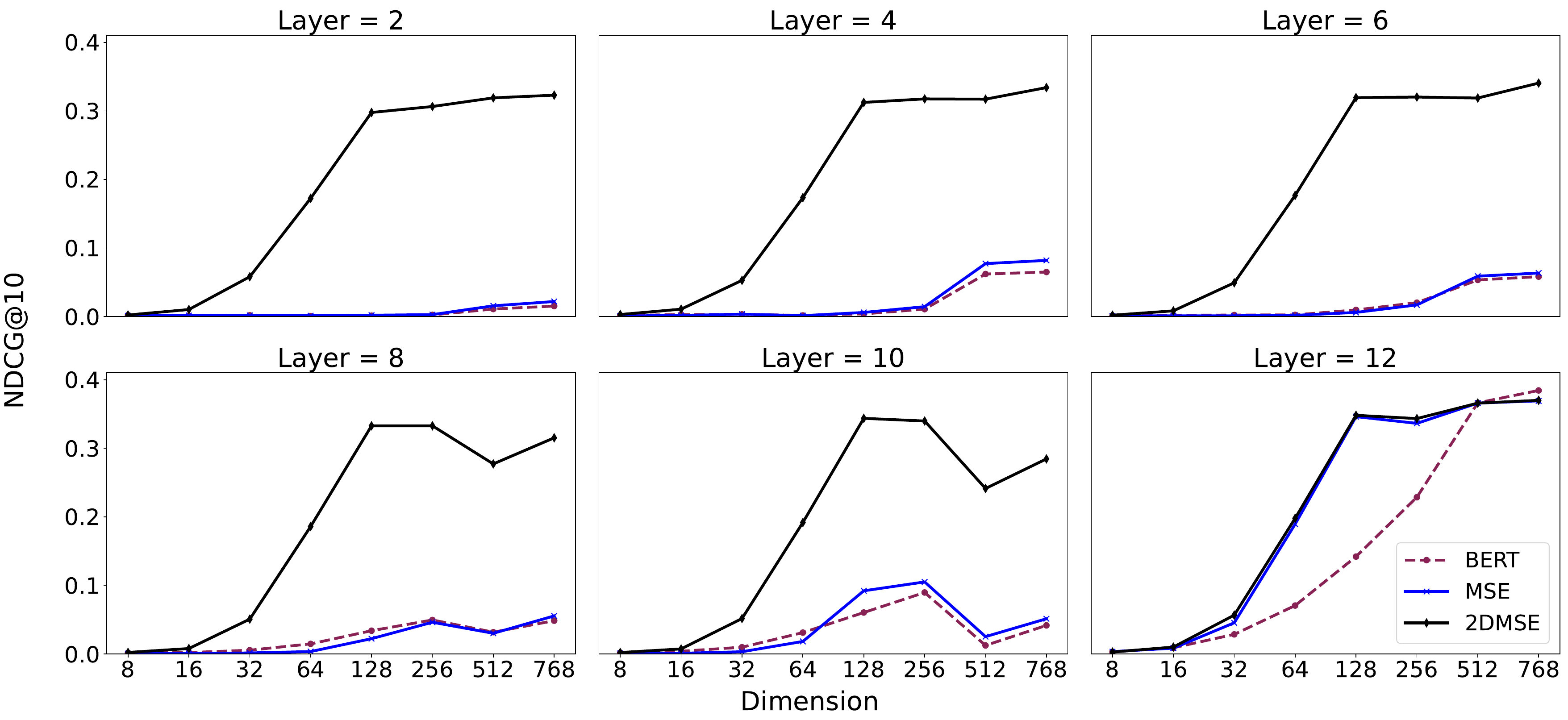}
	}
	\caption{Comparison of Model Effectiveness in different layer and dimension setup for retrieval, evaluated using NDCG@10 averaged on BEIR datasets.}
	\label{fig:retreival_beir}
\end{figure}

Figure~\ref{fig:retreival_base} compares the retrieval effectiveness of 2DMSE and the baseline methods; note that we only use the implementation of the second version of 2DMSE as it has shown to generate better embeddings in STS tasks (see RQ1 results). 


Our analysis reveals that the performance trends observed in STS tasks generally extend to retrieval tasks. Specifically, 2DMSE outperforms MSE and BERT across all models except in the last layer, but it is less effective than BERT-S across all layer-dimension setups. A notable difference is that for STS tasks, small dimension embeddings lead to lower but still acceptable performance -- and performance degradation is smooth as embeddings get smaller. Conversely, in the retrieval task, performance abruptly deteriorateeas for dimensions lower than 128. At the same time, it may be surprising at first to observe that dimensions higher than 128 are also (often) deteriorating retrieval effectiveness, though not to the extent seen for $<$128. 
These declines may be attributed to two primary factors:
\begin{enumerate}
	\item The training and loss computation is tailored exclusively to the target dimension of 128 when sub-layers are employed, as specified in Equation~\ref{equ:10}; all other dimensions are not trained. While this does not influence the STS task, for retrieval this may be important for effectiveness.
	\item Retrieval tasks may inherently be more challenging than STS tasks, potentially requiring a more refined approach to loss computation.
\end{enumerate}

We also conducted experiments on BEIR datasets and found similar results as in MSMARCO, as demonstrated in Figure~\ref{fig:retreival_beir}.
These findings suggest a need for adjusting the training process of 2DMSE to retrieval tasks. We investigate how to improve retrieval effectiveness in subsequent sections.

%

\subsection{\textbf{RQ3: Does a modified loss calculation in 2DMSE improve retrieval effectiveness?}}

\label{sec:rq3}
\begin{figure}[t]
	\centering
	\resizebox{\textwidth}{!}{
		\includegraphics[width=\textwidth]{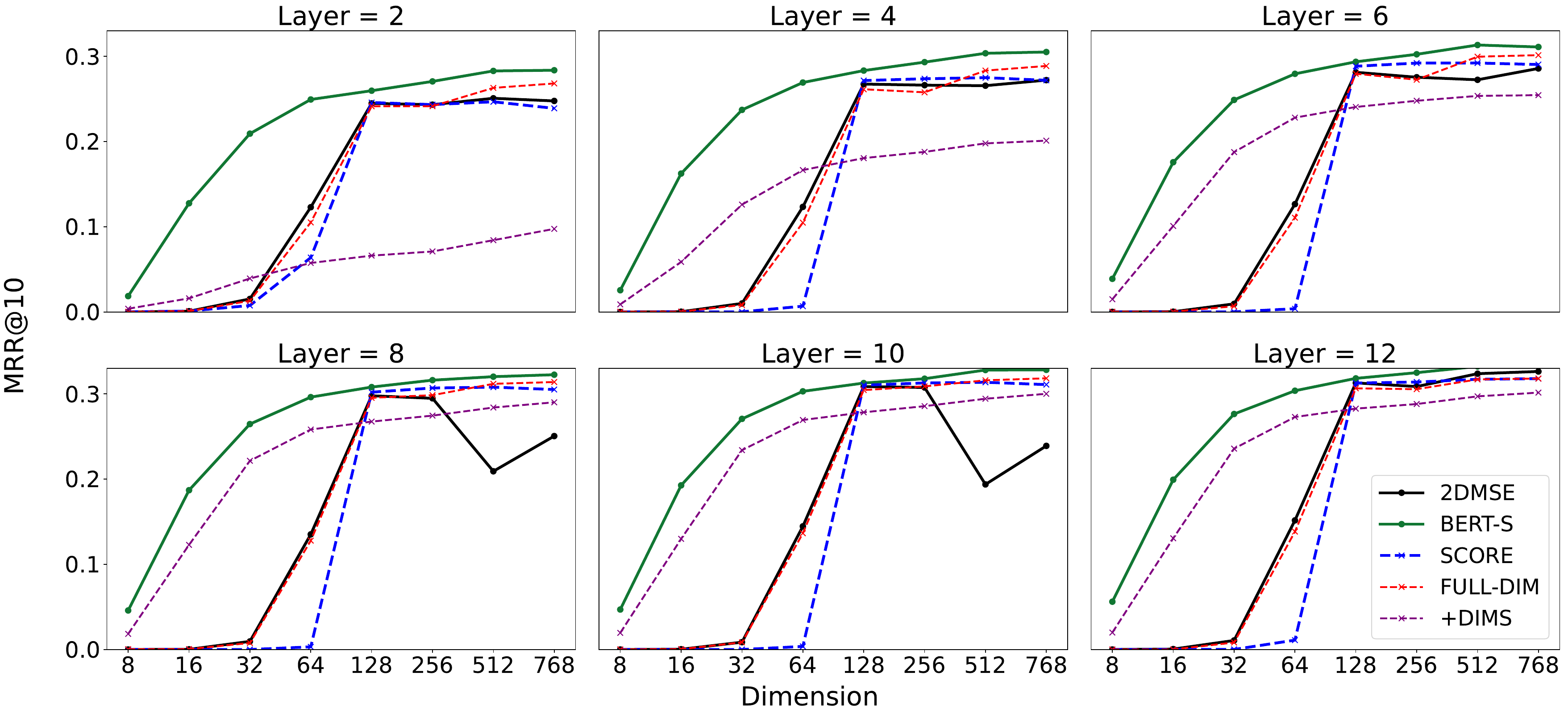}
	}
	\caption{Comparison of model effectiveness with different loss computations, evaluated using MRR@10 on MSMARCO.}
	\label{fig:retrieval_ablation}
\end{figure}

We explore three potential modifications to the 2DMSE model aimed at enhancing retrieval effectiveness: (1) integration of scoring-focused loss, (2) incorporation of full dimension loss from each layer, and (3) training on more target dimensions. The results demonstrating the impact of these modifications on model effectiveness are reported in Figure~\ref{fig:retrieval_ablation}.

\subsubsection{Scoring-Focused Loss (SCORE).}
\label{subsec:score_loss}
The current dimension-specific loss in 2DMSE combines KL divergence and MSE, aiming to minimize the distance between truncated embeddings and their PCA representations. By integrating a scoring-focused loss, we shift the objective to align more directly with the primary goal of effective document ranking. This modification involves using relevance scores to assess the effectiveness of smaller versus full dimensions. Initial findings suggest that while SCORE helps maintain effectiveness at higher dimensions, it does not improve generalization at lower dimensions as effectively as traditional embedding-based methods.

\subsubsection{Full Dimension Loss from Each Layer (FULL-DIM).}
\label{subsec:full_dim_loss}
Originally, 2DMSE focused only on a single target sub-dimension from all layers. In this modification, we expand this to include full dimensions from each layer, hypothesizing that leveraging the complete informational capacity of each layer would enhance the model's learning capabilities. Our results show effectiveness comparable to the original 2DMSE for dimensions  $<$128, with significant improvement for dimensions $\ge$128. Notably, this modification resolved the drop in performance observed of the original 2DMSE at layers 8 and 10 for dimension 512.

\subsubsection{Training on More Target Dimensions (+DIMS).}
\label{subsec:more_dims}
Our analysis, depicted in Figure~\ref{fig:retreival_base}, demonstrates that the effectiveness diminishes when the dimension is less the the target dimension 128. To address this, we expanded the training objectives to cover a broader range of dimensions\footnote{We only show results that with training on target dimensions [8, 16, 32, 64, 128 256, 512]; more results at ~\href{https://github.com/ielab/2DMSE-Reproduce/}{Github Repository}}. From the results, we find that adding more target dimensions smoothed the effectiveness curve, making it a similar shape to the training outcomes of small-scale, separately trained BERT models. However, this was at the expense of reduced performance at higher dimensions. Moreover, the more dimensions we include in the training objective, the more smoothed impact this has, i.e. higher effectiveness on lower dimensions, and lower on higher dimensions.


\subsection{\textbf{RQ4: Can a small, 2DMSE query encoder be used in conjunction with a full document encoder to improve effectiveness and efficiency in retrieval?}}

\begin{figure}[t]
	\centering
	\resizebox{\textwidth}{!}{
		\includegraphics[width=\textwidth]{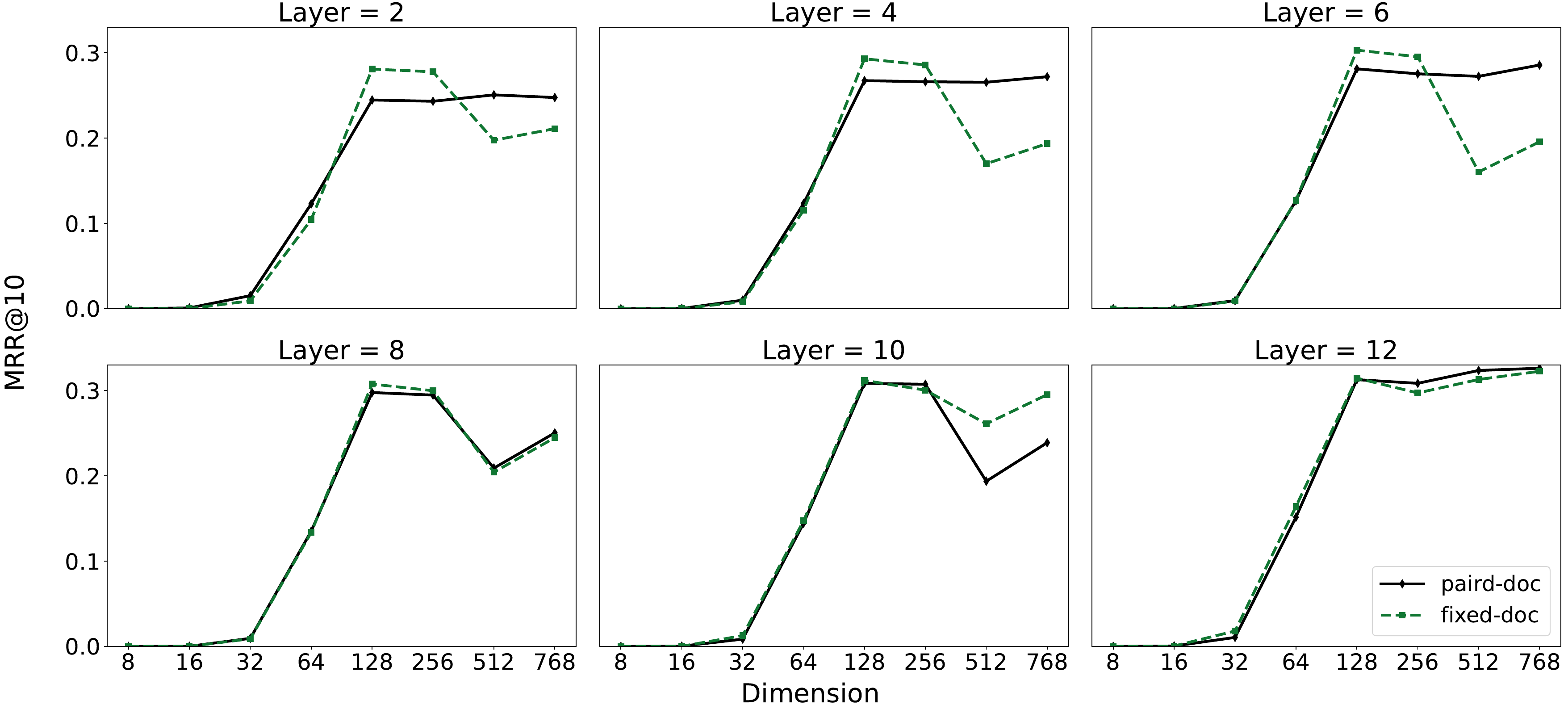}
	}
	\caption{Comparison of model effectiveness when a different document encoder is used, evaluated using MRR@10 on MSMARCO.}
	\label{fig:retreival_ablation_2}
	
\end{figure}

To explore the effectiveness of using a fixed full-size document encoder in combination with a scalable, smaller query encoder, we conducted an experiment with a modified 2DMSE model. This model adopts two distinct training strategies: (1) fixing the document encoder to use only the output from the last layer of the model when computing loss (we label this as \texttt{fix-doc}), and (2) excluding dimension losses for passages from all sub-layers. We then compared the effectiveness of \texttt{fix-doc} and of the original 2DMSE (\texttt{paired-doc}) across all layer and dimension setups. In \texttt{fix-doc}, the document encoder is consistently fixed to the full model output.

The results, illustrated in Figure~\ref{fig:retreival_ablation_2}, indicate that using a fixed document encoder does not consistently benefit all layer and dimension setups. Specifically, it significantly underperforms compared to a paired-encoder approach when small layer query encoders (i.e., layers 2, 4, 6) are used for large dimension setups. However, it provides higher effectiveness when using large layer query encoders (i.e., layers 8, 10). Despite this variability, the use of a fixed full-size document encoder leads to increased document indexing computations compared to paired encoders of smaller sizes, which may not be justified given the inconsistent performance improvements.


%
%

\section{Related Works}
Transformer-based embedding models have demonstrated effectiveness across various NLP tasks, including Sentence Text Similarity (STS)~\cite{williams-etal-2018-broad} and Dense Retrieval~\cite{gao-callan-2021-condenser,karpukhin-etal-2020-dense,zhuang2024promptrepspromptinglargelanguage,wang2023balanced}. The STS task evaluates the capability of embeddings to measure the semantic closeness between two sentences, with a higher score indicating a closer semantic relationship. Dense retrieval tasks assess the effectiveness based on the ability to retrieve relevant passages to a query by mapping queries and documents into embeddings and computing similarities. Both tasks share a common challenge: efficient encoding, particularly as model sizes have grown exponentially~\cite{li2024eseespressosentenceembeddings}. This growth has resulted in increased computational costs associated with transformer-based embedding models.

To address the high computational demands of encoding in large transformer models,  approaches have been proposed to develop pruned models that could speed up text encoding. The KALE method involves training a pruned query encoder using only the initial sub-layers of a full model, significantly reducing the number of parameters involved in query encoding~\cite{campos-etal-2023-quick}. The encoder is then fine-tuned by minimizing the KL divergence between its query representations and those from the final layer of a fully trained model. The MatFormer method, rather than eliminating layers, reduces the dimensionality of the model to achieve higher efficiency~\cite{devvrit2023matformer}. Similarly, Matryoshka training~\cite{mrl2022} focuses on dimensionality reduction, but it specifically targets the resulting embeddings, maintaining the original encoding cost while enhancing efficiency during similarity scoring. Building on this, the 2D Matryoshka training approach utilizes both a smaller embedding size and embeddings derived from sub-layers, reducing latency in encoding (by using sub-layers) and enhancing efficiency in similarity scoring (by reducing dimensionality)~\cite{li2024eseespressosentenceembeddings}.

\section{Conclusion}

We have successfully reproduced the 2D Matryoshka training approach, initially proposed for the Sentence Text Similarity task, and extended its application to retrieval tasks. Our reproduction efforts addressed and resolved discrepancies between two different implementations of the original 2D Matryoshka model, confirming the consistency of the reported trends, except that 2DMSE-V2 can not achieve a higher effectiveness than separately trained small-scale models. We also find that 2DMSE-V2 showcases a better design choice than 2DMSE-V1, demonstrated by a higher effectiveness across most sub-size models.

Our investigation confirms that the 2D Matryoshka training approach generalizes to retrieval tasks. Experiments with different loss strategies showed improvements in the original 2DMSE model for retrieval tasks. By incorporating full-dimension loss and extending training to more dimensions, we enhanced overall model performance and mitigated the reduced effectiveness at lower dimensions. However, we find that contrary to expectations, fixing the document encoder to only utilize the last layer's output did not yield substantial benefits compared to configurations where both documents and queries are encoded using the same layers. 

We recognize several limitations in our current work: (1) We used the same layer weighting strategy for loss computation as in the original work, without exploring the impact of different weighting strategies on retrieval tasks; and (2) Our investigation was limited to the bert-base-uncased model, without considering other potentially more effective decoder-based models. The constraints of our experiments, which already required over 600 GPU hours on H100 GPUs\footnote{This figure does not account for initial trials that resulted in errors.}, precluded further exploration in these areas. These points are outlined as directions for future work.


\bibliographystyle{splncs04}
\bibliography{reference}

\appendix
%
%
%
%
%

\end{document}